\documentclass[12pt]{article}
\usepackage[english]{babel}%
\usepackage{graphicx}
\usepackage{amsfonts}

\begin{document}

\newtheorem{theorem}{Theorem}
\newtheorem{cor}{Corollary}
\newtheorem{example}{Example}

\renewcommand{\thesection}{\arabic{section}}
\renewcommand{\thetheorem}{\arabic{section}.\arabic{theorem}}
\renewcommand{\theequation}{\arabic{section}.\arabic{equation}}
\renewcommand{\theexample}{\arabic{section}.\arabic{example}}
\renewcommand{\thecor}{\arabic{section}.\arabic{cor}}

\def\qed{{\bf \hfill $\Box$}\endtrivlist}
\def\endsolve{{\bf \hfill $\bigtriangleup$}\endtrivlist}

\begin{center}{\Large Some Analytics for Steiner Minimal Tree Problem for Four Terminals} \end{center}

\begin{center}{\large Alexei Yu. Uteshev\footnote{alexeiuteshev@gmail.com}}  \end{center}

\begin{center}{\it St. Petersburg State University, St. Petersburg, Russia} \end{center}


Given the coordinates of four terminals  in the Euclidean plane we present explicit formulas for Steiner point coordinates for Steiner minimal tree problem. We utilize the obtained formulas for evaluation of the influence of terminal coordinates on the loci of Steiner points.


\section{Introduction}\label{SIntro}
\setcounter{equation}{0}
\setcounter{theorem}{0}
\setcounter{example}{0}

\textbf{Problem.} Given set of points $ \mathbb P= \{P_j\}_{j=1}^n $ in the Euclidean plane, find a system of line segments such that their union   forms a connected set $ \mathbb U $ containing $ \mathbb P $, and such that the total length of the line segments is minimized.

This problem of construction the shortest possible network interconnecting the points of the set $ \mathbb P $ is known as the (\textbf{Euclidean})  \textbf{Steiner minimal tree problem} (\textbf{SMT} problem).

The stated problem in its particular case of $ n=3 $ noncollinear points is known since 1643 as (classical) Fermat-Torricelli problem.
It has a unique solution which
\begin{itemize}
\item either coincides with the system of three segments connecting the points $ P_1,P_2,P_3 $ with the so-called \emph{Fermat} or \emph{Fermat-Torricelli point}  of the triangle $ P_1P_2P_3 $; this is the case where every angle of the triangle $ P_1P_2P_3 $ is less than $ 2\pi/3 $;
\item or, in case where there exists a vertex $ P_j $ with the triangle angle equal to or greater than $ 2\pi/3 $, it coincides with the system of two triangle sides meeting at  $ P_j $.
\end{itemize}

For the sake of concordance with the foregoing definition, we will refer to the Fermat-Torricelli point of the triangle as to its \textbf{Steiner point}.

The case of $ n>3 $ points is much harder in treatment. First of all, the Sreiner minimal tree problem should be distinguished from another problem frequently called  the \emph{Fermat-Torricelli problem}, the latter consists in finding a \underline{single} junction point $ S $ which minimize $ \sum_{j=1}^n |SP_j| $. However the solution for this problem even for the case of $ n=4 $ points located at the vertices of a square does not provide the shortest network. It turns out that two junction points are necessary \cite{Gilbert&Pollak}. Thus, for the general case, the problem is reduced to determining of the set\footnote{Possibly empty!} $ \mathbb S=\{S_1,\dots,S_k\} $ of $ k\ge 0 $ extra junction points and the line segments connecting them with the points of the set $ \mathbb P $. We refer to \cite{Brazil_2014} for a comprehensive and intriguing historical review of the theory.

 The general results for the problem are formulated in terms of Graph Theory \cite{Cock70}. A tree $ \mathbb U $ with vertices  $ \mathbb P \cup \mathbb S  $ and rectilinear edges linking certain pairs of vertices is a \textbf{Steiner tree on} $ \mathbb P $ iff it has the following properties
\begin{itemize}
  \item[\textbf{(P1)}] $ \mathbb U $ is non-self-intersecting.
  \item[\textbf{(P2)}] The \emph{valency}\footnote{Or \emph{degree}} of every $ S_i $ equals $ 3 $.
  \item[\textbf{(P3)}] The valency of every $ P_j $ is $ \le 3 $.
  \item[\textbf{(P4)}] Each $ S_j $ is the Steiner point of the triangle formed by the points which directly connect $ S_j $ in $ \mathbb U $.
  \item[\textbf{(P5)}] $ 0 \le k \le n-2 $.
\end{itemize}
The points $ P_j $ are usually called \textbf{terminals} while the \emph{junction} points $ S_i $ are referred to as \textbf{Steiner points} of a Steiner tree.

For any given  $ \mathbb P $, there are finitely many Steiner trees and at least one of them is the Steiner minimal  tree \cite{b1}.
A \textbf{full Steiner tree} on $ \mathbb P $ is a tree which satisfies \textbf{(P1)}-\textbf{(P4)} and also has $ k=n-2 $. Any Steiner minimal  tree is a union of full Steiner subtrees.

The present paper is devoted to the four terminals SMT problem. We restrict ourselves here with the cases of full Steiner trees, i.e. the trees with exactly two Steiner points. We aim at finding the conditions for existence of such trees as well as the coordinates of Steiner points; both problems are to be solved \underline{algebraically} in terms of terminal coordinates.
The paper can be treated as a continuation of the previous work \cite{Uteshev14} on the generalized Fermat-Torricelli (three terminals) problem.

In comparison with the admirable elegance of geometric approaches for the problem, the representation of its solution in final analytical form (Section \ref{Stein4}) looks like prosaic, cumbersome and boring. The only arguments which might excuse the present author for their creation are the following:
\begin{itemize}
\item these formulas are universal and yield the exact result (i.e., free of truncation errors);
\item they provide one with an opportunity to evaluate the influence of parameters involved in the problem on its solution.
\end{itemize}

The latter problem is briefly touched in Section \ref{Scor} where we investigate the dynamics of Steiner points under the variation of a certain terminal.

\textbf{Notation.} We denote the coordinates of the terminal $ P_j $ by $ (x_j,y_j) $,
$$ r_{jk} = |P_jP_k|= \sqrt{(x_j-x_k)^2+(y_j-y_k)^2} \ ; $$
and by
$$ \langle \overrightarrow{A_1A_2}, \overrightarrow{A_3A_4} \rangle $$
the inner product of the vectors.

For the terminals $ \{P_j\}_{j=1}^4 $ and Steiner points $ S_1, S_2 $, the representation
$$ \begin{array}{c} P_1 \\ P_2 \end{array} S_1 S_2 \begin{array}{c} P_4 \\ P_3 \end{array} $$
has the meaning that terminals $ P_1 $ and $ P_2 $ are both connected to Steiner point $ S_1 $, that $ P_3 $ and $ P_4 $ are both connected to Steiner point $ S_2 $; finally $ S_1 $ and $ S_2 $ are connected with each other. We will refer to such a representation as to \textbf{topology of Steiner tree}.

\section{Three terminals} \label{Stein3}
\setcounter{equation}{0}
\setcounter{theorem}{0}
\setcounter{example}{0}

We first recall the geometrical construction of Steiner point for three terminals $ P_1, P_2 $ and $ P_3 $ . The algorithm outlined in the following example is a combination of Torricelli and Simpson approaches for the construction of Fermat-Torricelli point for the triangle $P_1P_2P_3 $.

\begin{example} \label{Ex0}  For the terminals
$$
P_1=(4,4), \ P_2= (2,1), P_4=(7,1)
$$
construct the Steiner minimal tree.
\end{example}

\textbf{Solution.} First construct the equilateral triangle on the segment $ P_1P_2 $  and outside the triangle $ P_1P_2P_3 $ (Fig. 1).
\begin{figure}
\begin{center}
\includegraphics[width=0.50\textwidth]{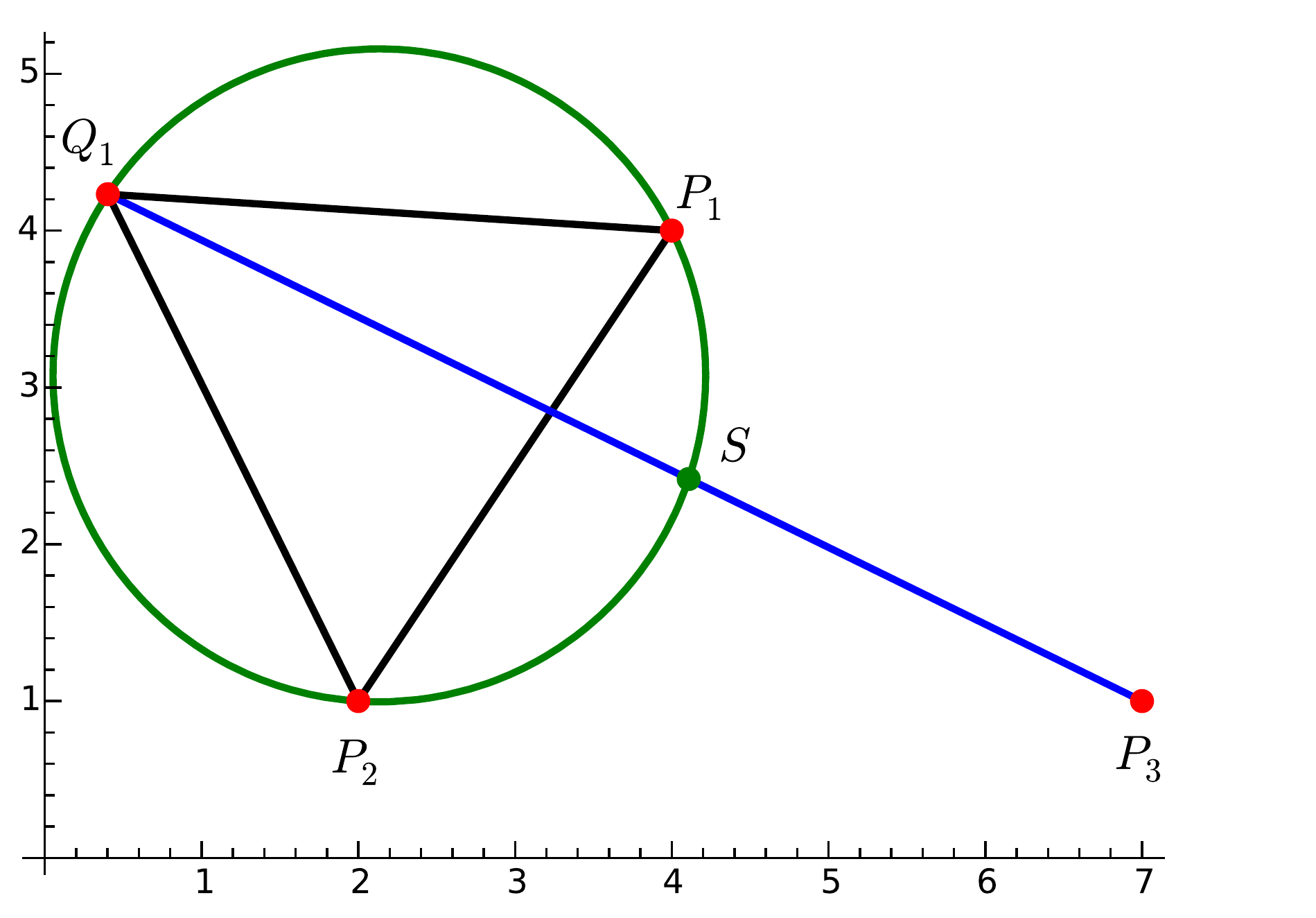}
\caption{}
\end{center}
\end{figure}
Denote its third vertex by $ Q_1 $. Next, draw the circle circumscribing this triangle; we denote it as $ \mathfrak{C} $. Finally draw the segment through $ Q_1 $ and $ P_3 $.
\begin{figure}
\begin{center}
\includegraphics[width=0.50\textwidth]{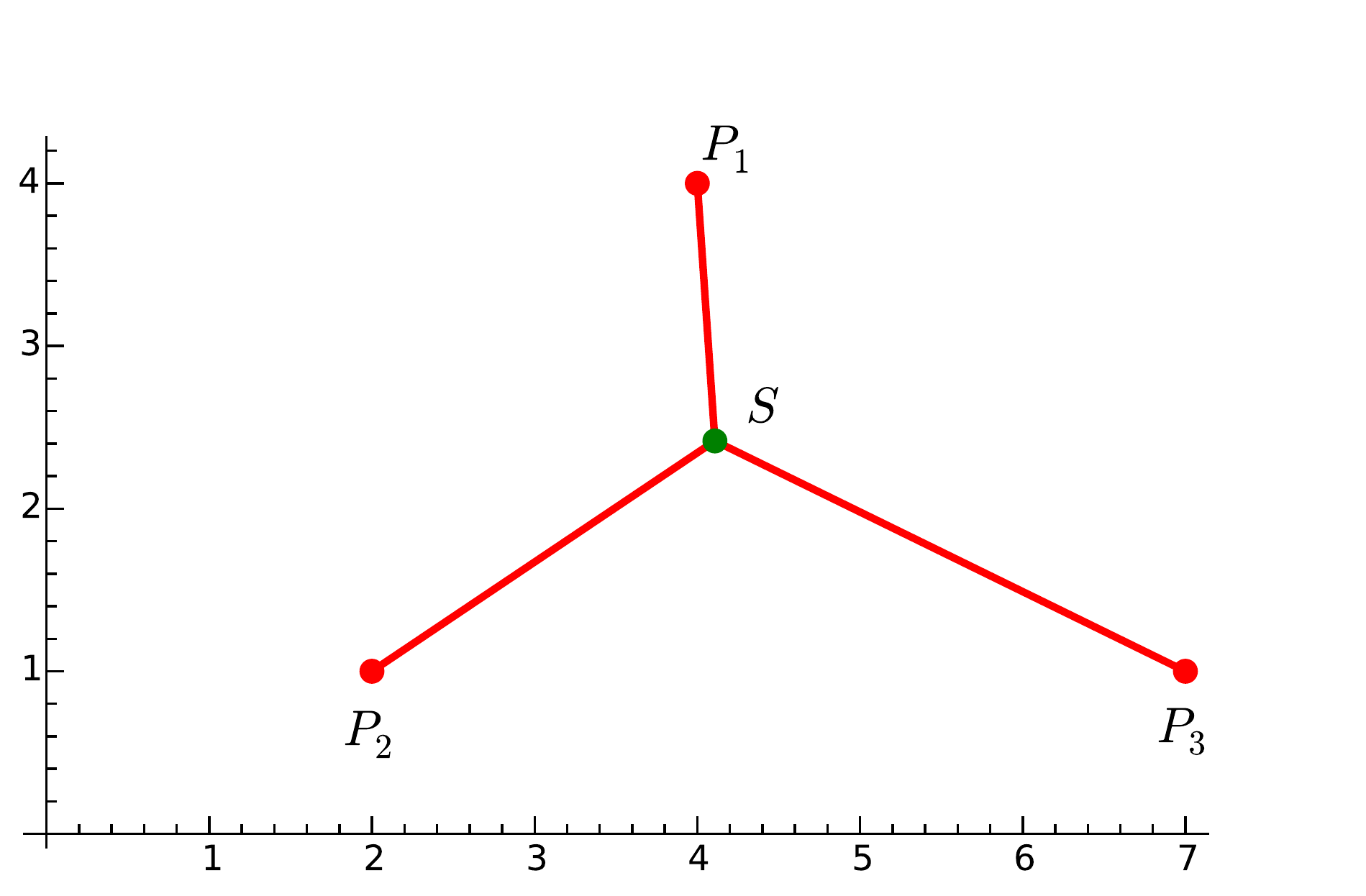}
\caption{Steiner tree for three terminals}
\end{center}
\end{figure}
The intersection point of the segment with the circle $ \mathfrak{C} $ is Steiner points for the tree (Fig. 2).
The length of SMT equals $|Q_1P_3| $. \endsolve

Analytical solution for the SMT problem is given by the following result\footnote{It is presented here with some changes in notation: in \cite{Uteshev14} $ d $ denotes \underline{squared} length and $ S $ stands for $ \mathfrak S $!} \cite{Uteshev14}:

\begin{theorem} \label{Steiner3} Let all the angles of the triangle $ P_1P_2P_3 $ be less than $ 2\pi/3 $, or, equivalently:
$$ r_{12}^2+r_{13}^2+r_{12}r_{13}-r_{23}^2>0,\  r_{23}^2+r_{12}^2+r_{12}r_{23}-r_{13}^2>0,\  r_{13}^2+r_{23}^2+r_{13}r_{23}-r_{12}^2>0 \ . $$
The coordinates of Steiner point for the terminals $ P_1,P_2,P_3 $ are as follows:
\begin{equation}
x_{\ast}=\frac{\kappa_1\kappa_2\kappa_3}{2 \sqrt{3} |\mathfrak S|  d^2} \left(\frac{x_1}{\kappa_1}+\frac{x_2}{\kappa_2}+\frac{x_3}{\kappa_3} \right), \
y_{\ast}=\frac{\kappa_1\kappa_2\kappa_3}{2 \sqrt{3} |\mathfrak S| d^2} \left(\frac{y_1}{\kappa_1}+\frac{y_2}{\kappa_2}+\frac{y_3}{\kappa_3} \right)
\label{FTp}
\end{equation}
with the length of the SMT equal to
\begin{equation}
d=\sqrt{\frac{\kappa_1+\kappa_2+\kappa_3}{\sqrt{3}}}=\sqrt{\frac{r_{12}^2+r_{13}^2+r_{23}^2}{2}+ \sqrt{3}\, |\mathfrak S|} .
\label{Stein-dist3}
\end{equation}
Here
\begin{equation}
\mathfrak S=\left|\begin{array}{ccc} 1 & 1 & 1 \\ x_1 & x_2 & x_3 \\ y_1 & y_2 & y_3  \end{array} \right|  \, ,
\label{S}
\end{equation}
(and therefore $ | \mathfrak S| $ equals the doubled area of the triangle $ P_1P_2P_3 $), while
$$
\kappa_1=\frac{\sqrt{3}}{2}(r_{12}^2+r_{13}^2-r_{23}^2)+|\mathfrak S| , \ \kappa_2=\frac{\sqrt{3}}{2}(r_{23}^2+r_{12}^2-r_{13}^2)+|\mathfrak S| , \
\kappa_3=\frac{\sqrt{3}}{2}(r_{13}^2+r_{23}^2-r_{12}^2)+|\mathfrak S | .
$$
\end{theorem}

\begin{cor}\label{cor1}  Let the loci of terminals $ P_1 $ and $ P_2 $ be fixed in the $ (x,y) $-plane while the locus of the terminal $ P_3 $ be variable with the restrictions that the points $ P_1, P_2 $ and $ P_3 $ are numbered counterclockwise and satisfy the conditions of Theorem \ref{Steiner3}. The locus of Steiner point for terminals $ P_1,P_2,P_3 $ coincides with the arc of the circle $ \mathfrak{C} $ constructed via the algorithm outlined in solution of Example \ref{Ex0}. It has its center at
\begin{equation}
C=\left(\frac{1}{2}x_1+\frac{1}{2}x_2-\frac{1}{2\sqrt{3}}y_1+\frac{1}{2\sqrt{3}}y_2\, ,\ \frac{1}{2\sqrt{3}}x_1-\frac{1}{2\sqrt{3}}x_2+\frac{1}{2}y_1+\frac{1}{2}y_2 \right)
\label{circle_c}
\end{equation}
and its radius equal to $ r_{12}/\sqrt{3} $. This circle is the circumscribing one for the equilateral triangle $ P_1P_2Q_1 $ with
\begin{equation}
Q_1=\left(\frac{1}{2}x_1+\frac{1}{2}x_2-\frac{\sqrt{3}}{2}y_1+\frac{\sqrt{3}}{2}y_2\, ,\ \frac{\sqrt{3}}{2}x_1-\frac{\sqrt{3}}{2}x_2+\frac{1}{2}y_1+\frac{1}{2}y_2 \right) \, .
\label{Q1}
\end{equation}
\end{cor}

\section{Four terminals} \label{Stein4}
\setcounter{equation}{0}
\setcounter{theorem}{0}
\setcounter{example}{0}

We will start the treatment of this case with recalling the geometrical algorithm for Steiner tree construction \cite{Gilbert&Pollak}.

\begin{example} \label{Ex1}  For the terminals
$$
P_1=(2,6), \ P_2= (1,1), P_3=(9,2), P_4=(6,7)
$$
construct Steiner tree of the topology $ \begin{array}{c} P_1 \\ P_2 \end{array} S_1 S_2 \begin{array}{c} P_4 \\ P_3 \end{array} $.
\end{example}

\textbf{Solution.} First construct two equilateral triangles on the sides $ P_1P_2 $ and $ P_3P_4 $ and outside the quadrilateral $ P_1P_2P_3P_4 $ (Fig.3).
\begin{figure*}
\begin{center}
\includegraphics[width=0.70\textwidth]{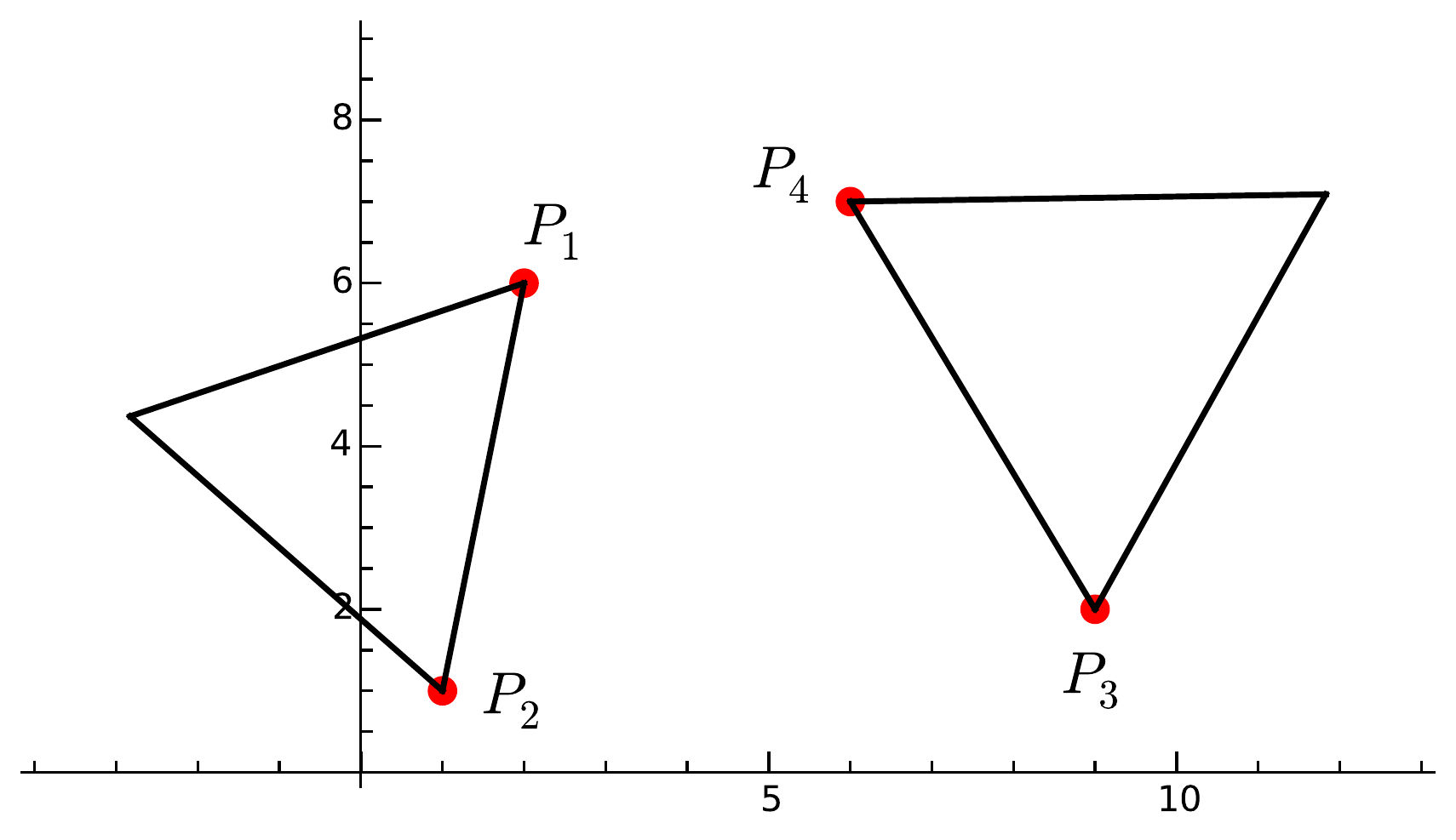}
\caption{}
\end{center}
\end{figure*}
Their third vertices are denoted by $ Q_1 $ and $ Q_2 $ correspondingly.
Next, draw two circles circumscribing these triangles (Fig.4).
\begin{figure*}
\begin{center}
\includegraphics[width=0.70\textwidth]{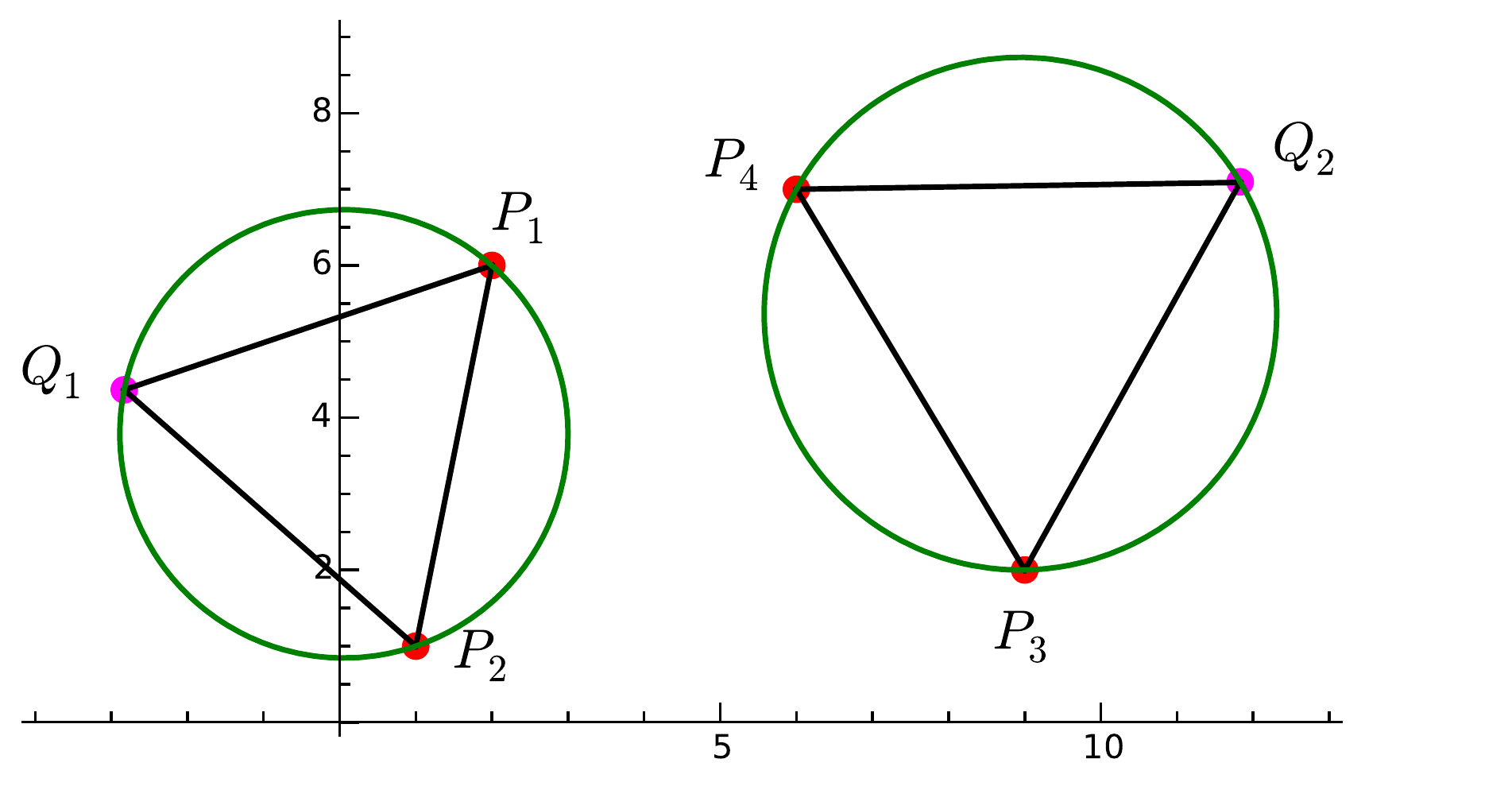}
\caption{}
\end{center}
\end{figure*}
Finally draw the segment through $ Q_1 $ and $ Q_2 $ (Fig.5).
\begin{figure*}
\begin{center}
\includegraphics[width=0.70\textwidth]{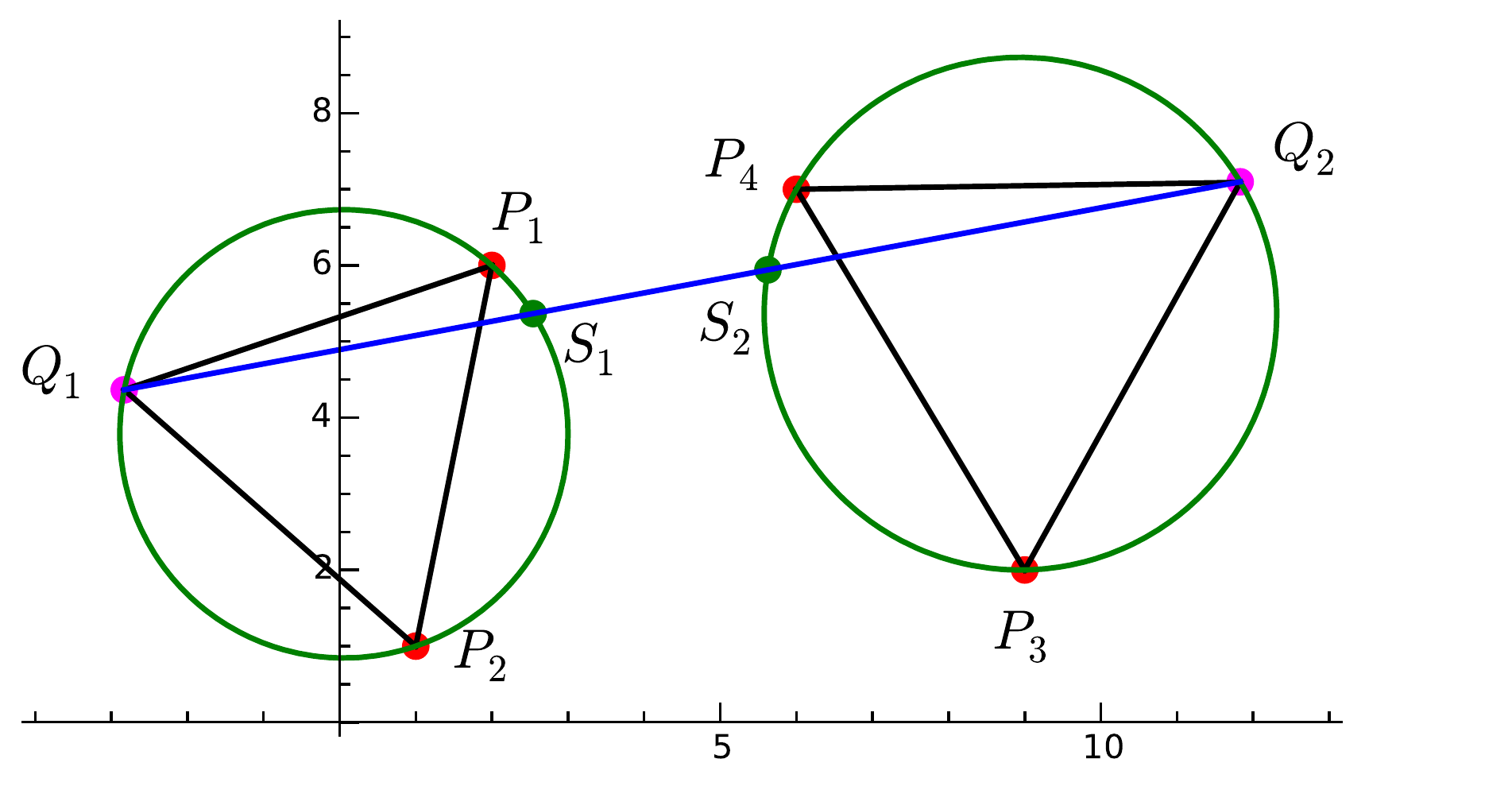}
\caption{}
\end{center}
\end{figure*}
The intersection points of the segment with the circles are Steiner points for the tree. The length of the obtained tree (Fig.6) equals $ |Q_1Q_2| $.
\begin{figure*}
\begin{center}
\includegraphics[width=0.70\textwidth]{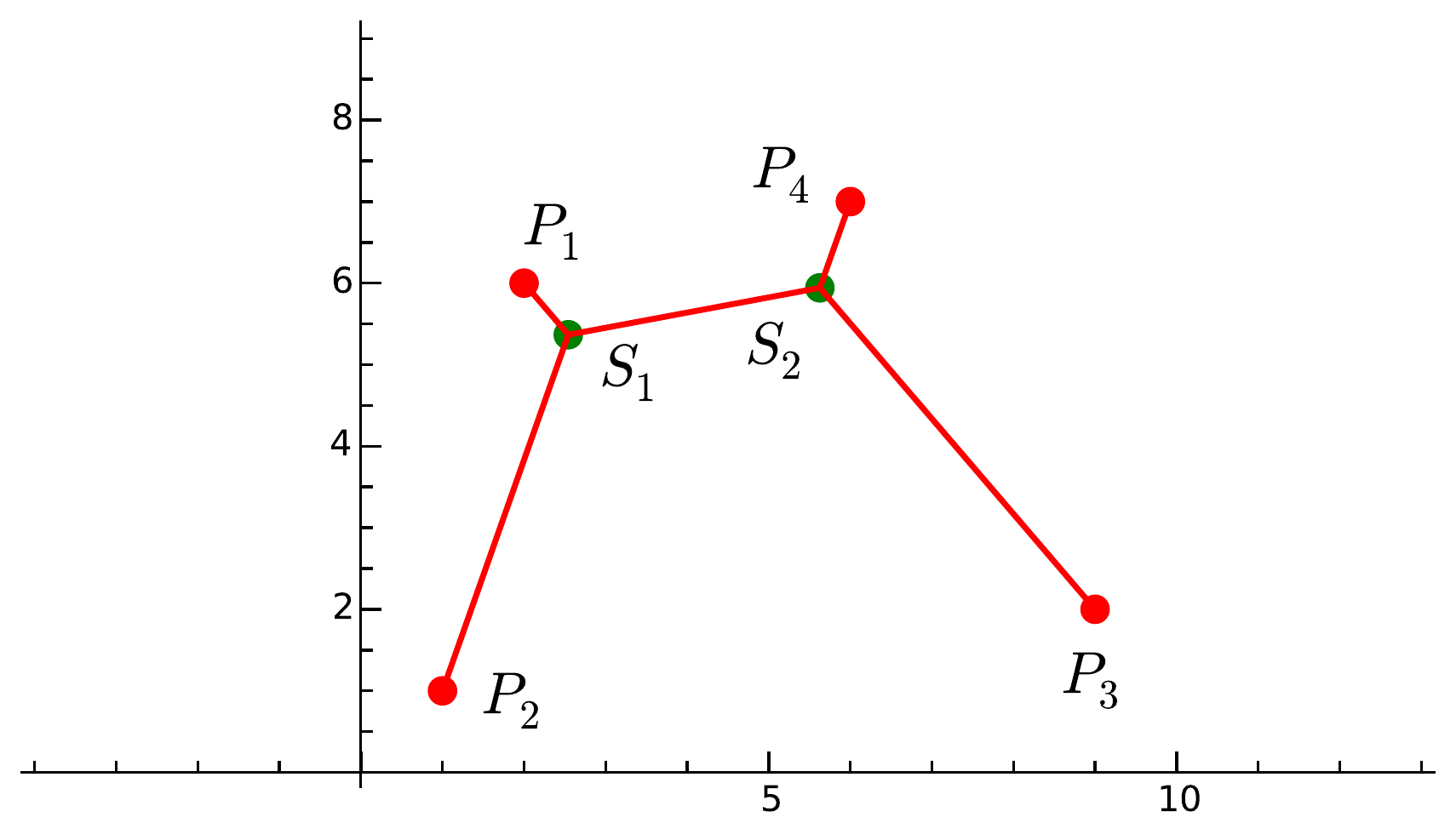}
\caption{Steiner tree for four terminals}
\end{center}
\end{figure*}

\endsolve

As for the general case, the problem of applicability of the suggested algorithm for Steiner tree construction is in question: one can easily imagine such a quadrilateral for which the algorithm results in points $ S_1 $ and $ S_2 $ located \emph{outside} it; thus they do not satisfy the condition \textbf{(P4)} mentioned in Section \ref{SIntro}.

\textbf{Assumption.} Hereinafter we will treat the case where the terminals $ \{P_j\}_{j=1}^4 $, while counted counterclockwise,  compose a convex quadrilateral $P_1P_2P_3P_4$.

The condition of convexity is a necessary one for the existence of a full Steiner tree for the given set of terminals \cite{Pollak}.

\begin{theorem} \label{Steiner4} Let the preceding assumption be fulfilled.
Set
\begin{eqnarray}
\tau_1 &= &  2\, x_1 - x_2 -2\, x_3 + x_4 +\sqrt{3} (y_2- y_4), \label{tau1} \\
\tau_2 & = &  - x_1 +2\, x_2 + x_3 -2\, x_4 +\sqrt{3} (y_3- y_1), \label{tau2}
\end{eqnarray}
and
\begin{equation}
\eta_1=-\frac{1}{\sqrt{3}}(\tau_1+2\,\tau_2),\ \eta_2 = \frac{1}{\sqrt{3}}(2\,\tau_1+\tau_2)  \label{eta12} \ .
\end{equation}
If all the values
\begin{equation}
\delta = -(x_1-x_3) \eta_1 + (y_1-y_3) \tau_1 ,
\label{delta}
\end{equation}
\begin{equation}
\left\{\begin{array}{cc}
\begin{array}{c}
\delta_1 = (x_1-x_2) \eta_2 - (y_1-y_2) \tau_2,  \\
\delta_2 = (x_1-x_2) \eta_1 - (y_1-y_2) \tau_1,
\end{array}
&
\begin{array}{c}
\delta_3 = -(x_3-x_4) \eta_2 + (y_3-y_4) \tau_2,  \\
\delta_4 = -(x_3-x_4) \eta_1 + (y_3-y_4) \tau_1
\end{array}
\end{array}
\right.
\label{delta1234}
\end{equation}
are positive then there exists a Steiner tree of the topology $ \begin{array}{c} P_1 \\ P_2 \end{array} S_1 S_2 \begin{array}{c} P_4 \\ P_3 \end{array} $.
The coordinates of Steiner point $ S_1 $ are as follows:
\begin{equation}
x_{\ast}=x_1 - \frac{\sqrt{3}}{2} \cdot \frac{\delta_1 \tau_1}{\tau_1^2+\tau_1 \tau_2+ \tau_2^2}, \quad y_{\ast}=y_1 - \frac{\sqrt{3}}{2} \cdot \frac{\delta_1 \eta_1}{\tau_1^2+\tau_1 \tau_2+ \tau_2^2}
\label{Steiner pt1}
\end{equation}
and those of $ S_2 $:
\begin{equation}
x_{\ast \ast}=x_3 + \frac{\sqrt{3}}{2} \cdot \frac{\delta_3 \tau_1}{\tau_1^2+\tau_1 \tau_2+ \tau_2^2}, \quad
y_{\ast \ast}=y_3 + \frac{\sqrt{3}}{2} \cdot \frac{\delta_3 \eta_1}{\tau_1^2+\tau_1 \tau_2+ \tau_2^2} \ .
\label{Steiner pt2}
\end{equation}
The length of the tree  equals to
\begin{equation}
d= \sqrt{\frac{\tau_1^2+\tau_1 \tau_2+ \tau_2^2}{3}} \, . \label{Stein-dist4}
\end{equation}
\end{theorem}

\textbf{Proof.} Consider the objective function corresponding to the assumed topology of the tree
\begin{equation}
F(x_{\ast},y_{\ast},x_{\ast \ast},y_{\ast \ast})=|S_1P_1|+|S_1P_2|+|S_1S_2|+|S_2P_3|+|S_2P_4| \ .
\label{obj_fun}
\end{equation}
The system providing its stationary points is as follows
\begin{equation} \left\{
\begin{array}{ccc}
\displaystyle \frac{x_{\ast}-x_1}{|S_1P_1|}   + & \displaystyle \frac{x_{\ast}-x_2}{|S_1P_2|}  + & \displaystyle \frac{x_{\ast}-x_{\ast \ast}}{|S_1S_2|} = 0, \\
\displaystyle \frac{y_{\ast}-y_1}{|S_1P_1|} + & \displaystyle \frac{y_{\ast}-y_2}{|S_1P_2|} +  & \displaystyle \frac{y_{\ast}-y_{\ast \ast}}{|S_1S_2|} = 0, \\
\displaystyle \frac{x_{\ast \ast}-x_3}{|S_2P_3|}   + & \displaystyle \frac{x_{\ast \ast}-x_4}{|S_2P_4|}  + & \displaystyle \frac{x_{\ast \ast}-x_{\ast}}{|S_1S_2|} = 0, \\
\displaystyle \frac{y_{\ast \ast}-y_3}{|S_2P_3|}   + & \displaystyle \frac{y_{\ast \ast}-y_4}{|S_2P_4|}  + & \displaystyle \frac{y_{\ast \ast}-y_{\ast}}{|S_1S_2|} = 0.
\end{array}
\right.
\label{stat_points}
\end{equation}
To prove the first of these formulas  for the values of $ x_{\ast},y_{\ast}, x_{\ast \ast},y_{\ast \ast} $ given by (\ref{Steiner pt1}) and (\ref{Steiner pt2}), we will first deduce the alternative representation for these values:
\begin{equation}
x_{\ast}=x_2 - \frac{\sqrt{3}}{2} \cdot \frac{\delta_2 \tau_2}{\tau_1^2+\tau_1 \tau_2+ \tau_2^2}, \quad y_{\ast}=y_2 - \frac{\sqrt{3}}{2} \cdot \frac{\delta_2 \eta_2}{\tau_1^2+\tau_1 \tau_2+ \tau_2^2} \, .
\label{Steiner pt10}
\end{equation}
The equivalence of the first formula to its counterpart from (\ref{Steiner pt1}) stems from
$$
\delta_1 \tau_1 - \delta_2 \tau_2 \stackrel{(\ref{delta1234})}{=} \left[(x_1-x_2) \eta_2 - (y_1-y_2) \tau_2\right] \tau_1 -
\left[(x_1-x_2) \eta_1 - (y_1-y_2) \tau_1\right] \tau_2
$$
$$
=(x_1-x_2)\left[\eta_2 \tau_1-\eta_1 \tau_2\right] \stackrel{(\ref{eta12})}{=} (x_1-x_2) \left[ \frac{1}{\sqrt{3}}(2\,\tau_1+\tau_2) \tau_1 +\frac{1}{\sqrt{3}}(\tau_1+2\,\tau_2) \tau_2 \right]
$$
$$
= \frac{2}{\sqrt{3}} (x_1-x_2)  \left[\tau_1^2+ \tau_1 \tau_2 + \tau_2^2 \right] \ .
$$
The second formula from (\ref{Steiner pt10}) can be proven similarly.

Next, with the aid of representations (\ref{Steiner pt1}) and (\ref{Steiner pt10}), one can prove that
$$
|S_1P_1|
$$
$$
= \sqrt{(x_{\ast}-x_1)^2+(y_{\ast}-y_1)^2}\stackrel{(\ref{Steiner pt1})}{=}\sqrt{\frac{3}{4} \cdot \frac{\delta_1^2 \tau_1^2 + \delta_1^2 \eta_1^2}{(\tau_1^2+ \tau_1 \tau_2 + \tau_2^2)^2}} \stackrel{(\ref{eta12})}{=} \sqrt{\frac{\delta_1^2}{\tau_1^2+ \tau_1 \tau_2 + \tau_2^2}} $$
\begin{equation}
= \frac{\delta_1}{\sqrt{\tau_1^2+ \tau_1 \tau_2 + \tau_2^2}}
\label{S1P1}
\end{equation}
(the last equality follows from the assumption $ \delta_1 > 0 $ stated in the theorem) and, similarly, that
\begin{equation}
|S_1P_2|= \frac{\delta_2}{\sqrt{\tau_1^2+ \tau_1 \tau_2 + \tau_2^2}} \ .
\label{S1P2}
\end{equation}

Finally we intend to deduce the equalities
\begin{equation}
x_{\ast}-x_{\ast \ast}= \frac{\sqrt{3}}{2} \cdot \frac{(\tau_1+\tau_2) \delta}{\tau_1^2+ \tau_1 \tau_2 + \tau_2^2}, \quad
y_{\ast}-y_{\ast \ast}= \frac{1}{2} \cdot \frac{ (\tau_1-\tau_2) \delta}{\tau_1^2+ \tau_1 \tau_2 + \tau_2^2}
\label{minus}
\end{equation}
from which evidently follows\footnote{Under the assumption $ \delta > 0 $ stated in the theorem.}
\begin{equation}
|S_1S_2|= \frac{\delta}{\sqrt{\tau_1^2+ \tau_1 \tau_2 + \tau_2^2}} \ .
\label{S1S2}
\end{equation}
The proof of the first of the formulas (\ref{minus})  is based on the equality
\begin{equation}
\delta_1+\delta_3 = (x_1-x_3)(\eta_1+\eta_2)-(y_1-y_3)(\tau_1+\tau_2)
\label{eq_w0}
\end{equation}
the validity of which can be established as follows:
$$
\delta_1+\delta_3 \stackrel{(\ref{delta1234})}{=} (x_1-x_2-x_3+x_4) \eta_2+(-y_1+y_2+y_3-y_4) \tau_2
$$
$$
=\frac{1}{\sqrt{3}}  (x_1-x_2-x_3+x_4) (2\tau_1+\tau_2) +(-y_1+y_2+y_3-y_4) \tau_2
$$
$$
=\frac{2}{\sqrt{3}}(x_1-x_2-x_3+x_4)\tau_1+\left[ \frac{1}{\sqrt{3}} (x_1-x_2-x_3+x_4)+(-y_1+y_2+y_3-y_4) \right] \tau_2
$$
$$
=\left[ \frac{1}{\sqrt{3}} (x_1-x_3)-y_1+y_3 \right] \tau_1 +\left[ \frac{1}{\sqrt{3}} (x_1-2\,x_2-x_3+2\,x_4)+y_1-y_3 \right] \tau_1
$$
$$
+
\left[ \frac{1}{\sqrt{3}} (2\,x_1-x_2-2\,x_3+x_4)+y_2-y_4 \right] \tau_2 + \left[ \frac{1}{\sqrt{3}} (-x_1+x_3)-y_1+y_3 \right]\tau_2
$$
$$
=\frac{1}{\sqrt{3}}(x_1-x_3)(\tau_1-\tau_2)-(y_1-y_3)(\tau_1+\tau_2)=(x_1-x_3)(\eta_1+\eta_2)-(y_1-y_3)(\tau_1+\tau_2) \, .
$$
We now start deducing the first formula from (\ref{minus}):
\begin{equation}
x_{\ast}-x_{\ast \ast} \stackrel{(\ref{Steiner pt1}), (\ref{Steiner pt2})}{=} x_1-x_3-
\frac{\sqrt{3}}{2} \cdot \frac{(\delta_1+\delta_3) \tau_1}{\tau_1^2+\tau_1 \tau_2+ \tau_2^2} \, .
\label{eq_w1}
\end{equation}
Let us manipulate with the numerator of the fraction from the last formula:
$$
(\delta_1+\delta_3)\tau_1 \stackrel{(\ref{eq_w0})}{=}\left[(x_1-x_3)(\eta_1+\eta_2)-(y_1-y_3)(\tau_1+\tau_2) \right]\tau_1
$$
$$
=\left[(x_1-x_3)(\eta_1+\eta_2)\tau_1-(x_1-x_3)(\tau_1+\tau_2) \eta_1 \right] +
\left[(x_1-x_3)(\tau_1+\tau_2)\eta_1-(y_1-y_3)(\tau_1+\tau_2) \tau_1 \right]
$$
$$
=(x_1-x_3)\left[ (\eta_1+\eta_2)\tau_1- (\tau_1+\tau_2)\eta_1 \right] +(\tau_1+\tau_2)\left[(x_1-x_3) \eta_1 - (y_1-y_3) \tau_1 \right]
$$
$$
\stackrel{(\ref{eta12}),(\ref{delta})}{=}(x_1-x_3)\left[ \frac{1}{\sqrt{3}}(\tau_1-\tau_2)\tau_1 + \frac{1}{\sqrt{3}}(\tau_1+\tau_2)(\tau_1+2\, \tau_2)
\right]- (\tau_1+\tau_2) \delta
 $$
$$
=\frac{2}{\sqrt{3}} (x_1-x_3) (\tau_1^2+ \tau_1 \tau_2 + \tau_2^2) - (\tau_1+\tau_2) \delta \ .
$$
Substitution of the last expression into the right-hand side of (\ref{eq_w1}) results in  the first formula from (\ref{minus}).
The second formula from (\ref{minus}) can be deduced similarly.

Applying now the formulas (\ref{Steiner pt1}) and (\ref{S1P1}) for the first summand from the left-hand side of the first formula from (\ref{stat_points}), the formulas (\ref{Steiner pt10}) and (\ref{S1P2}) for the second summand, (\ref{minus}) and (\ref{S1S2}) for the third summand, one gets the true equality. The validity of the remained equalities can be established similarly.

We collect now all the formulas for the segment lengths
\begin{equation}
 |P_1S_1|=\frac{\delta_1}{\sqrt{3} d},\ |P_2S_1|=\frac{\delta_2}{\sqrt{3} d},\ |P_3S_2|=\frac{\delta_3}{\sqrt{3} d},\ |P_4S_2|=\frac{\delta_4}{\sqrt{3} d},\ |S_1S_2|=\frac{\delta}{\sqrt{3} d}
\end{equation}
and use them for finding the critical value of the objective function (\ref{obj_fun}). One has
$$ \delta_1+\delta_2+\delta_3+\delta_4+ \delta $$
$$ \stackrel{(\ref{delta1234})}{=} (x_1-x_2-x_3+x_4) (\eta_1+\eta_2)+ (-y_1+y_2+y_3-y_4) (\tau_1+\tau_2) + \delta $$
$$=(x_1-x_3) \eta_1 + (-y_1+y_3) \tau_1 + (-x_2+x_4) \eta_1 + (x_1-x_2-x_3+x_4) \eta_2 + (y_2-y_4) \tau_1 +
(-y_1+y_2+y_3-y_4) \tau_2 + \delta $$
$$ \stackrel{(\ref{delta}),(\ref{eta12})}{=} -(-x_2+x_4) \frac{(\tau_1+2\tau_2)}{\sqrt{3}} + (x_1-x_2-x_3+x_4) \frac{(2\tau_1+\tau_2)}{\sqrt{3}} + (y_2-y_4) \tau_1 + (-y_1+y_2+y_3-y_4) \tau_2
$$
$$
=\frac{\tau_1}{\sqrt{3}} \left(2\, x_1-x_2-2\,x_3+x_4+\sqrt{3}(y_2-y_4) \right) + \frac{\tau_2}{\sqrt{3}} \left(
x_1+x_2-x_3-x_4+\sqrt{3}(-y_1+y_2+y_3-y_4) \right)
$$
$$
 \stackrel{(\ref{tau1}),(\ref{tau2})}{=} \frac{1}{\sqrt{3}} \tau_1^2 + \frac{1}{\sqrt{3}} \tau_2 (\tau_1+\tau_2) = \frac{1}{\sqrt{3}} (\tau_1^2+\tau_1 \tau_2 + \tau_2^2) \, .
$$
This yields the claimed critical value  (\ref{Stein-dist4}) for the objective function (\ref{obj_fun}).

We do not prove here that this value is indeed the minimal one. Instead of this we will discuss the meaning of the positivity restrictions imposed in the theorem on the values $ \delta, \delta_1,\dots,\delta_4 $. If we alter simultaneously their sign to negative then substitution of formulas (\ref{Steiner pt1}) and (\ref{Steiner pt2}) into (\ref{stat_points}) still keep them valid.  What reason underlines then the choice of the sign for deltas mentioned in the theorem? To answer this question let us determine the relative position of the points $ S_1, S_2  $ with respect to the quadrilateral $ P_1P_2P_3P_4 $. The validity of the following equalities:
$$
\left|
\begin{array}{lll}
1 & 1 & 1 \\
x_1 & x_2 & x_{\ast} \\
y_1 & y_2 & y_{\ast}
\end{array}
\right| = \frac{\sqrt{3}}{2} \cdot \frac{\delta_1 \delta_2 }{\tau_1^2+\tau_1 \tau_2 + \tau_2^2}\, , \quad
\left|
\begin{array}{lll}
1 & 1 & 1 \\
x_2 & x_3 & x_{\ast} \\
y_2 & y_3 & y_{\ast}
\end{array}
\right| = \frac{\sqrt{3}}{2} \cdot \frac{\delta \delta_2 + \delta_2 \delta_3}{\tau_1^2+\tau_1 \tau_2 + \tau_2^2}\, , \
$$
$$
\left|
\begin{array}{lll}
1 & 1 & 1 \\
x_3 & x_4 & x_{\ast} \\
y_3 & y_4 & y_{\ast}
\end{array}
\right| = \frac{\sqrt{3}}{2} \cdot \frac{\delta_3 \delta_4+ \delta_3 \delta+\delta_4 \delta}{\tau_1^2+\tau_1 \tau_2 + \tau_2^2}\, , \
\left|
\begin{array}{lll}
1 & 1 & 1 \\
x_4 & x_1 & x_{\ast} \\
y_4 & y_1 & y_{\ast}
\end{array}
\right|=\frac{\sqrt{3}}{2} \cdot \frac{\delta \delta_1 + \delta_1 \delta_4}{\tau_1^2+\tau_1 \tau_2 + \tau_2^2}
$$
can be verified  by direct substitution of the formulas for $ x_{\ast}, y_{\ast}, x_{\ast \ast}, y_{\ast \ast} $ and deltas from the statement of the theorem. Due to assumption of the theorem, all the left-hand side determinants are positive. Therefore the point $ S_1 $ lies inside the quadrilateral $ P_1P_2P_3P_4 $. Similar arguments lead to the same claim for the point $ S_2 $.

For the final check, let us evaluate the angles $ P_1S_1P_2 $ and $ P_1S_1S_2 $
$$
\frac{\left<\overrightarrow{S_1P_1}, \overrightarrow{S_1P_2}\right>}{|S_1P_1|\cdot |S_1P_2|}
=\frac{(x_{\ast}-x_1)(x_{\ast}-x_2)+(y_{\ast}-y_1)(y_{\ast}-y_2)}{|S_1P_1|\cdot |S_1P_2|}
\stackrel{(\ref{Steiner pt1}),(\ref{Steiner pt10}) \atop (\ref{S1P1}),(\ref{S1P2})}{=}
\frac{3/4(\tau_1 \tau_2+ \eta_1 \eta_2)}{\tau_1^2+\tau_1 \tau_2 + \tau_2^2} \stackrel{(\ref{eta12})}{=}- \frac{1}{2} \, ;
$$
$$
\frac{\left<\overrightarrow{S_1P_1}, \overrightarrow{S_1S_2}\right>}{|S_1P_1|\cdot |S_1S_2|}
\stackrel{(\ref{Steiner pt1}),(\ref{Steiner pt2}) \atop (\ref{S1P1}),(\ref{S1P2})}{=}\frac{-\frac{\sqrt{3}}{2}\left( \tau_1(x_1-x_3)+\eta_1(y_1-y_3)\right)+ \delta_1+ \delta_3}{\delta}
$$
$$
\stackrel{(\ref{delta}),(\ref{delta1234})}{=} \frac{-\frac{\sqrt{3}}{2}\left( \tau_1(x_1-x_3)+\eta_1(y_1-y_3)\right)- \delta+(x_1-x_3) \eta_2+(-y_1+y_3)\tau_2}{\delta}\stackrel{(\ref{eta12}),(\ref{delta})}{=}-\frac{1}{2}\, .
$$
Therefore both angles in question equals $ 2\pi/3 $. This agrees with the property \textbf{(P4)} of Steiner points mentioned in Section 1. \qed

\begin{example} \label{Ex2} Find the coordinates for Steiner points for topology from Example \ref{Ex1}.
\end{example}

\textbf{Solution.} Here
$$
\tau_1=-9-6\sqrt{3},\ \tau_2=-3-4\sqrt{3},\ \eta_1= 14+5\sqrt{3},\ \eta_2= -16-7\sqrt{3} \, .
$$
The conditions of Theorem \ref{Steiner4} are fulfilled: the values
$$
\delta=62+11\sqrt{3},\ \delta_1=-1+13\sqrt{3},\ \delta_2=59+35\sqrt{3},\ \delta_3=63+ 41\sqrt{3},\ \delta_4=3+ 15\sqrt{3}
$$
are positive. Further,
$$ \tau_1^2+\tau_1 \tau_2 + \tau_2^2 = 345+186\sqrt{3} $$
and therefore the length of Steiner tree equals
$$
d=\sqrt{115+62\sqrt{3}} \approx 14.912651 \, .
$$
Formulas (\ref{Steiner pt1}) and (\ref{Steiner pt2}) yield the coordinates of Steiner points
\begin{eqnarray*}
x_{\ast}&=& \displaystyle \frac{571+323\sqrt{3}}{2(115+62\sqrt{3})}=\frac{5587}{3386}+\frac{1743}{3386}\sqrt{3} \approx 2.541631, \\
y_{\ast}&=& \displaystyle \frac{3609+2051\sqrt{3}}{6(115+62\sqrt{3})}=\frac{11183}{3386}+\frac{12107}{10158}\sqrt{3} \approx 5.367094
\end{eqnarray*}
and
\begin{eqnarray*}
x_{\ast \ast}&=& \displaystyle \frac{3(441+227\sqrt{3})}{2(115+62\sqrt{3})}=\frac{25479}{3386}-\frac{3711}{3386}\sqrt{3} \approx 5.626509, \\
y_{\ast \ast}&=& \displaystyle \frac{1349+747\sqrt{3}}{2(115+62\sqrt{3})}= \frac{16193}{3386}+\frac{2267}{3386}\sqrt{3}\approx 5.941984.
\end{eqnarray*}
\endsolve

\textbf{Remark 1.} Comparing the formulas for Steiner points (\ref{Steiner pt1}) and (\ref{Steiner pt2}) with their counterparts  (\ref{FTp}) for the three terminals problem, one may watch the following property: The denominators of all the formulas for Steiner point coordinates contain an explicit expression for the length of the corresponding tree. It looks like every Steiner point ``knows'' the length of the tree which this point is a part of\footnote{The dependency of the numerators of the fractions (\ref{FTp}) from an extra parameter, namely the area of the triangle $ P_1P_2P_3 $, is not essential: one can also extract this factor  from the numerators of the formulas and therefore eliminate it from the denominator \cite{Uteshev14}.}.

\textbf{Remark 2.} The condition $ \delta \ne 0 $ with $ \delta $ defined by (\ref{delta}) guarantees that the Steiner points $ S_1 $ and $ S_2 $ do not coincide. The restriction $ \delta > 0 $ can be reformulated in terms of geometry of the quadrilateral $ P_1P_2P_3P_4 $.
Indeed,
$$ \delta \stackrel{(\ref{delta})}{=} -(x_1-x_3) \eta_1 + (y_1-y_3) \tau_1  $$
$$ \stackrel{(\ref{tau1}),(\ref{eta12})}{=}
-(x_1-x_3)(-\sqrt{3}x_2+\sqrt{3}x_4+y_2-y_4+2\,y_3-2\, y_1)+(y_1-y_3)(2\,x_1-x_2-2\,x_3+x_4+\sqrt{3}y_2-\sqrt{3}y_4)
$$
$$
=-(x_1-x_3)(-\sqrt{3}x_2+\sqrt{3}x_4+y_2-y_4)+(y_1-y_3)(-x_2+x_4+\sqrt{3}y_2-\sqrt{3}y_4)
$$
$$
=2(x_3-x_1,y_3-y_1) \left(\begin{array}{rr} \sqrt{3}/2 & 1/2 \\ -1/2 &  \sqrt{3}/2 \end{array} \right) \left(\begin{array}{c} x_4-x_2 \\ y_4-y_2 \end{array} \right) \, .
$$
This value is positive iff the angle between the diagonal $ \overrightarrow{P_1P_3} $ of the quadrilateral and the other diagonal $ \overrightarrow{P_2P_4} $ turned through by $ \pi/6 $ clockwise is acute. Equivalently, if we denote by $ \psi $ the angle between diagonal vectors $ \overrightarrow{P_1P_3} $ and $ \overrightarrow{P_2P_4} $ then $ \delta $ is positive iff $ \psi < \pi/2 +\pi/6=2\, \pi/3 $.
This confirms the known condition for the existence of a full Sreiner tree \cite{Gilbert&Pollak}.

\section{Corollaries}\label{Scor}
\setcounter{equation}{0}
\setcounter{theorem}{0}
\setcounter{example}{0}
\setcounter{cor}{0}

In the present section some statements are presented without proofs: their  computer verification has been performed using the formulas from the previous section.

\subsection{Length of a tree}

We first transform the formula (\ref{Stein-dist4})  for the Steiner tree length:
\begin{equation}
d=\frac{1}{2}\sqrt{\frac{1}{3}(\tau_1-\tau_2)^2+(\tau_1+\tau_2)^2} =\frac{1}{2}\sqrt{A^2+B^2}
\label{dist_alt}
\end{equation}
for
\begin{eqnarray*}
A&=&\frac{1}{\sqrt{3}} (\tau_1-\tau_2)= \sqrt{3}(x_1-x_2-x_3+x_4)+(y_1+y_2-y_3-y_4), \\
B&=&\tau_1+\tau_2=(x_1+x_2-x_3-x_4)+\sqrt{3}(-y_1+y_2+y_3-y_4)\, .
\end{eqnarray*}
\textbf{Remark 3.}  This formula can be additionally checked with the aid of geometric algorithm of  Steiner tree construction  outlined  in  Example \ref{Ex1}.
Indeed, for the general case, the coordinates of the counterpart of the point  $ Q_1 $  from that construction are given by (\ref{Q1}), while those for $ Q_2 $ can be obtained from the latter via substitution $ 1 \rightarrow 3, 2 \rightarrow 4 $ for terminal coordinate subscripts:
$$
Q_2=\left(\frac{1}{2}x_3+\frac{1}{2}x_4-\frac{\sqrt{3}}{2}y_3+\frac{\sqrt{3}}{2}y_4\,,\ \frac{\sqrt{3}}{2}x_3-\frac{\sqrt{3}}{2}x_4+\frac{1}{2}y_3+\frac{1}{2}y_4  \right) \, .
$$
One can easily verify that  $ |Q_1 Q_2| $ equals (\ref{dist_alt}).

On expanding the radicand of (\ref{dist_alt}) further one gets
$$
d^2=(x_1-x_3)^2+(y_1-y_3)^2+(x_2-x_4)^2+(y_2-y_4)^2
$$
$$
-\left[ (x_1-x_3)(x_2-x_4)+(y_1-y_3)(y_2-y_4) \right]+ \sqrt{3} \left[(x_1-x_3)(y_2-y_4)-(y_1-y_3)(x_2-x_4) \right]
$$
$$
=r_{13}^2+r_{24}^2-r_{13}r_{24} \cos \psi + \sqrt{3} r_{13}r_{24} \sin \psi=r_{13}^2+r_{24}^2+2\, r_{13}r_{24} \cos \left(\frac{2\pi}{3}-\psi \right)
$$
where $ \psi $ denotes the angle between diagonal vectors $ \overrightarrow{P_1P_3} $ and $ \overrightarrow{P_2P_4} $. This result coincides with the one presented in \cite{Booth}.
Transforming it further to
$$= r_{13}^2+r_{24}^2-2\, r_{13}r_{24} \cos \left(\psi+ \frac{\pi}{3} \right) $$
we are able to provide it with the following geometric meaning which follows from the law of cosines:

\begin{cor}
The length of Steiner tree of the topology $ \begin{array}{c} P_1 \\ P_2 \end{array} S_1 S_2 \begin{array}{c} P_4 \\ P_3 \end{array} $ equals the length of the third side of the triangle constructed on two other sides coinciding with the diagonals of the quadrilateral $ P_1P_2P_3P_4 $ and with the angle between them chosen to be equal to $ \displaystyle \psi+ \pi/3 $.
\end{cor}

For the case of four terminals, there might exist a topology of full Steiner tree alternative to the one given in Theorem \ref{Steiner4}, namely $ \begin{array}{c} P_4 \\ P_1 \end{array} \tilde S_1 \tilde S_2 \begin{array}{c} P_3 \\ P_2 \end{array} $. To obtain the condition for its existence and the coordinates of Steiner points $ \tilde S_1 $ and $ \tilde S_2 $, one should make
the cyclic substitution
\begin{equation}
\left(
\begin{array}{cccc}
1 & 2 & 3 & 4 \\
2 & 3 & 4 & 1
\end{array}
\right)
\label{substitut}
 \ .
\end{equation}
for the terminal subscripts in the formulas of that theorem.

\begin{cor}\label{cor2} The necessary condition for the
existence of Steiner trees of both topologies  $ \begin{array}{c} P_1 \\ P_2 \end{array} S_1 S_2 \begin{array}{c} P_4 \\ P_3 \end{array} $ and $ \begin{array}{c} P_4 \\ P_1 \end{array} \tilde S_1 \tilde S_2 \begin{array}{c} P_3 \\ P_2 \end{array} $  is that the angles between the diagonals of the quadrilateral $ P_1P_2P_3P_4 $  are less than $ 2\pi/3 $ . If these trees exist with their lengths equal to correspondingly $ d $ and $ \tilde d $ then
$$ d^2-\tilde d^2= -2\left\{(x_3-x_1)(x_4-x_2)+(y_3-y_1)(y_4-y_2) \right\}=-2 \left<\overrightarrow{P_1P_3}, \overrightarrow{P_2P_4}\right> \, . $$
This means: If the diagonals of the quadrilateral are normal then both topologies give the minimal tree. Otherwise
the SMT coincides with the tree with Steiner points lying inside the acute vertical angles formed by the quadrilateral diagonals.
\end{cor}

\begin{example} \label{Ex3} For terminals from Example \ref{Ex1}, find Steiner points for the topology   $ \begin{array}{c} P_4 \\ P_1 \end{array} \tilde S_1 \tilde S_2 \begin{array}{c} P_3 \\ P_2 \end{array} $ .
\end{example}

\textbf{Solution.} The conditions of Theorem \ref{Steiner4} are fulfilled. One has
$$ \tilde S_1 = \left( \frac{34837}{7237}-\frac{14353}{43422}\sqrt{3} \ , \frac{21203}{14474}+\frac{28342}{21711}\sqrt{3} \right)
\approx (4.241211, 3.725958) \, ,
$$
$$
\tilde S_2 = \left(\frac{29648}{7237}-\frac{1109}{14474}\sqrt{3} \ , \ \frac{94173}{14474}-\frac{5092}{7237}\sqrt{3} \right)
\approx (3.964015, 5.287674 ) \, .
$$
and
$$ \tilde d= \sqrt{137+62\sqrt{3}} \approx 15.632887 \, . $$
Thus, Steiner tree of the topology $ \begin{array}{c} P_1 \\ P_2 \end{array} S_1 S_2 \begin{array}{c} P_4 \\ P_3 \end{array} $ drawn in Fig.6 and with coordinates for Steiner points presented in Example \ref{Ex2} is the SMT for the given set of terminals.
\endsolve

\begin{example} For terminals
$$
P_1=(1,6), \ P_2= (2,1 ), P_3=(6,1), P_4=(8,7) \, ,
$$
find Steiner points for the topologies  $ \begin{array}{c} P_1 \\ P_2 \end{array} S_1 S_2 \begin{array}{c} P_4 \\ P_3 \end{array} $ and $ \begin{array}{c} P_4 \\ P_1 \end{array} \tilde S_1 \tilde S_2 \begin{array}{c} P_3 \\ P_2 \end{array} $.
\end{example}

\textbf{Solution.} Here $ \left<\overrightarrow{P_1P_3},\overrightarrow{P_2P_4} \right>=0 $.  For the topology $ \begin{array}{c} P_1 \\ P_2 \end{array} S_1 S_2 \begin{array}{c} P_4 \\ P_3 \end{array} $  one has
\begin{eqnarray*}
S_1&=& \displaystyle \left(\frac{5997+ 1687\sqrt{3}}{3063},\ \frac{13014- 3103\sqrt{3}}{3063} \right) \approx (2.911841,\ 2.494106), \\
S_2& =& \displaystyle \left(\frac{7114}{1021}-\frac{3098}{3063}\sqrt{3},\  \frac{2973}{1021}-\frac{838}{3063}\sqrt{3}\right) \approx
(5.215836,\ 2.437983)
\end{eqnarray*}
while for $ \begin{array}{c} P_4 \\ P_1 \end{array} \tilde S_1 \tilde S_2 \begin{array}{c} P_3 \\ P_2 \end{array} $ one has
\begin{eqnarray*}
\tilde S_1& =& \displaystyle \left(\frac{3379}{1021}+\frac{923}{3063}\sqrt{3},\ \frac{6378}{1021}-\frac{3167}{3063}\sqrt{3} \right) \approx
(3.831434,\ 4.455956), \\
\tilde S_2&=& \displaystyle \left(\frac{4744}{1021}-\frac{1342}{3063}\sqrt{3},\ \frac{1263}{1021}+ \frac{1618}{3063}\sqrt{3} \right) \approx (3.887557,\ 2.151962) \, .
\end{eqnarray*}
The common value for tree length for these topologies equals
$$ \sqrt{122+60\sqrt{3}} \approx 15.030737 \, . $$
\endsolve

\textbf{Remark.} If the diagonals of $ P_1P_2P_3P_4 $ are normal then $ |S_1S_2| = |\tilde S_1\tilde S_2| $.

\subsection{Wandering terminal}

\begin{theorem}\label{th5}  Let the loci of terminals $ P_1,P_2 $ and $ P_4 $ be fixed in the $(x,y)$-plane while the locus of the terminal $ P_3 $ be variable with the only restriction that the coordinates of terminals $ \{ P_j\}_{j=1}^4 $ meet the conditions of Theorem \ref{Steiner4}.  For the topology $ \begin{array}{c} P_1 \\ P_2 \end{array} S_1 S_2 \begin{array}{c} P_4 \\ P_3 \end{array} $, the locus of  $ S_1 $ coincides with the arc of the circle $ \mathfrak{C} $  mentioned in Corollary \ref{cor1} while the locus of  $ S_2 $
coincides with the arc of the circle $ \widehat{\mathfrak{C}} $ with its center at
\begin{equation}
\widehat C= \left(\frac{1}{2}x_1+\frac{1}{2}x_4+\frac{1}{2\sqrt{3}}(-y_1+2\,y_2-y_4),\  \frac{1}{2}y_1+\frac{1}{2}y_4+\frac{1}{2\sqrt{3}}(x_1-2\,x_2+x_4) \right)
\label{hatC}
\end{equation}
and its radius equal to
$$
\widehat r_c =\sqrt{  \frac{1}{3} \left(\frac{r_{12}^2+r_{14}^2+r_{24}^2}{2}+\sqrt{3} \mathfrak S_{124} \right) } \, .
$$
Here
$$
\mathfrak S_{124}=\left|\begin{array}{ccc} 1 & 1 & 1 \\ x_1 & x_2 & x_4 \\ y_1 & y_2 & y_4  \end{array} \right|  \, ,
$$
(and therefore $ \mathfrak S_{124}  $ equals the doubled area of the triangle $ P_1P_2P_4 $).
\end{theorem}

\begin{cor} The circle $ \widehat{\mathfrak{C}} $ is the circumscribing circle for the equilateral triangle constructed on the side $ Q_1 P_4 $ with point  $ Q_1 $  introduced in Section \ref{Stein3} (i.e. it denotes the third vertex of the equilateral triangle built on the segment $ P_1 P_2 $ with its coordinates given by (\ref{Q1})). The circle $ \widehat{\mathfrak{C}} $ passes through the point
$$
I=\left(x_1+ \frac{\mathfrak S_{124}}{\sqrt{3}r_{24}^2} (\sqrt{3}(y_4-y_2)+x_2-x_4) ,\  y_1+ \frac{\mathfrak S_{124}}{\sqrt{3}r_{24}^2} (\sqrt{3}(x_2-x_4)+y_2-y_4)  \right)\, .
$$
of intersection of the circle $ \mathfrak{C} $ with the diagonal $ P_2P_4 $.
\end{cor}

\begin{example} For the terminals
$$ P_1=(5,8),P_2=(1,1),P_5=(10,7) $$
and for $ P_3 $ moving towards $ P_2 $ from the starting position at $ (11,3) $, the loci of Steiner points for the topology $ \begin{array}{c} P_1 \\ P_2 \end{array} S_1 S_2 \begin{array}{c} P_4 \\ P_3 \end{array} $ are displayed in Fig.7.
\begin{figure}
\begin{center}
\includegraphics[width=0.70\textwidth]{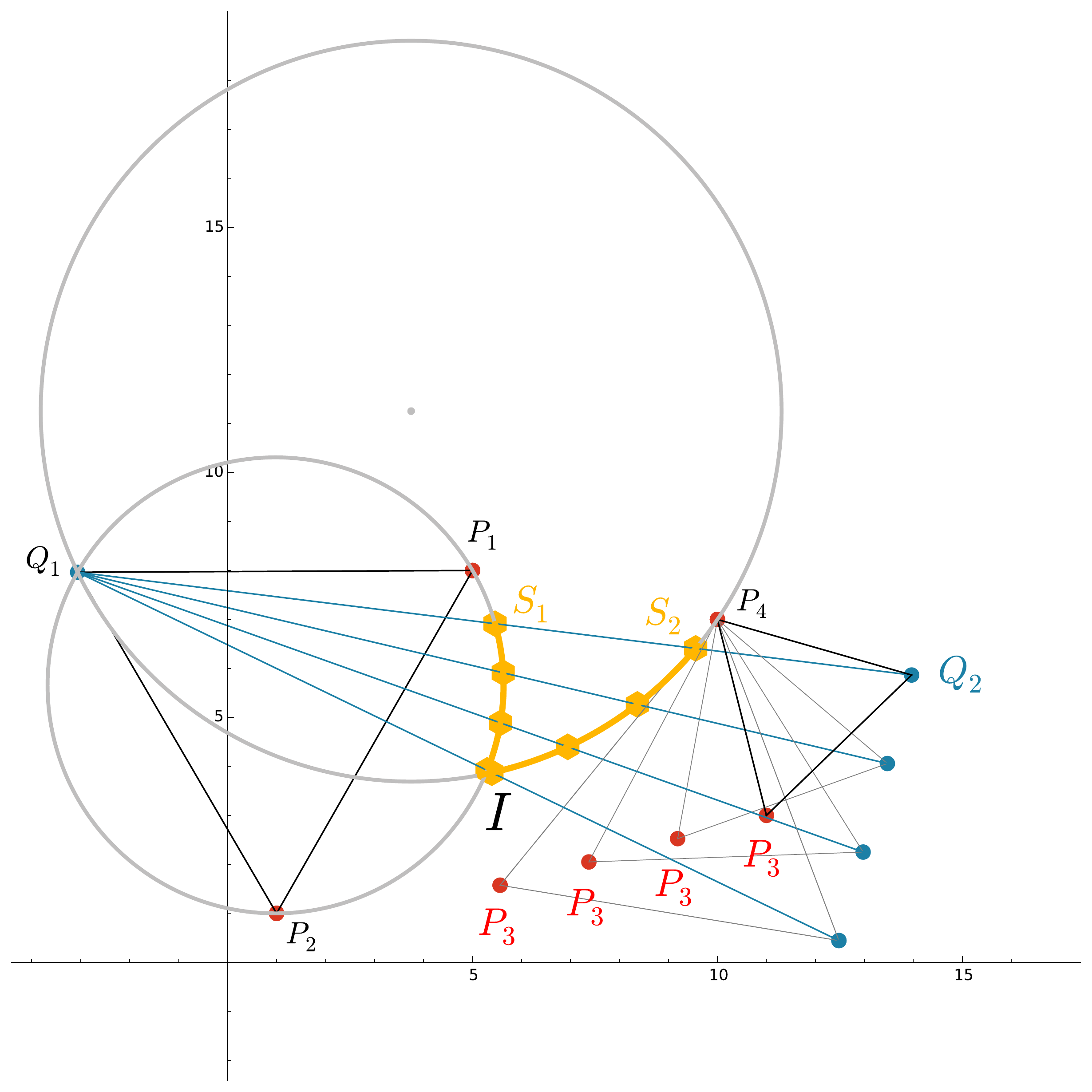}
\caption{Steiner points dynamics under variation of terminal $ P_3 $.}
\end{center}
\end{figure}
It should be noted that the trajectory of $ P_3 $ does not influence the trajectories of $ S_1 $ and $ S_2 $. The latter move along the corresponding arcs $ \mathfrak{C} $  and $ \widehat{\mathfrak{C}} $ mentioned in Theorem \ref{th5}  until $ P_3 $ meets the line $ P_1I $.
\end{example}

\section{Conclusions}

We presented some computational formulas for the Steiner minimal tree problem. The obtained solution is not complete since we have restricted ourselves with the case of full Steiner trees. It is also of potential interest to find analytical solution for  $ n=5 $ terminals --- just to satisfy the author's curiosity whether the observation mentioned in Remark 1 keeps to be fulfilled...

\vspace{1em}

{\bf Acknowledgments.} The author appreciate the courtesy of Elisabeth Semenova for designing Fig. 7.
This research was  supported by the St.Petersburg State University research grant \break \textbf{9.38.674.2013}.

\setcounter{equation}{0}
\setcounter{theorem}{0}
\setcounter{example}{0}

\end{document}